# Discovery of Graphene Sheets and C-Rich Micro-Oval structure in Stingless Bee Hive; Leading to an Emergent Material with Debut of Blue Emission

Manas Kumar Dalai[1,2,*], Ankita Mahakhuda[1] and Abinash Prusty[1]

[1]Mineralogy and Materials Characterization Department, CSIR – Institute of Minerals and Materials Technology, Bhubaneswar-751013, Odisha, India

[2]Academy of Scientific and Innovative Research (AcSIR), Ghaziabad - 201002, India

[*]Corresponding author: *dalaimk.immt@csir.res.in*

## ABSTRACT

Naturally produced stingless bee hive (NP-SBH) is an intricately produced material by the combination of waxes, resin and other biological materials that offers protection and structural stability to the bee colony. This study explores a detailed analysis of Indian stingless bee hive material using multi-characterization techniques approach to evaluate their morphological, ultrastructural, chemical composition and their crystallinity. FESEM reveal uniformly distributed micro-oval structures along with graphene sheets throughout the observed region. Furthermore, Energy Dispersive X-ray Analysis (EDAX) provides the richness of carbon (C) in graphene as well as in the micro-oval structure. HRTEM gives an insight about the internal ultrastructure and arrangement of atoms in the sample which revealed the presence of multiple graphene sheets. The ring shape electron diffraction pattern and high resolution lattice fringes provide the arrangement of carbon atoms, with interlayer spacing (d) value 3.4 Å, well agreed with that of graphene. Furthermore, X-ray Diffraction (XRD) and Fourier Transform Infrared (FTIR) spectroscopy support the presence of graphene. As a debut, we observe blue emission from PL spectroscopy with decay times 1.18 ns (42 %) and 5.41 ns (58 %).

## INTRODUCTION

Nature has its own way to create diverse, complex materials through natural processes to maintain ecological balance and environmental sustainability [1-5]. Nature acts as a circular system; where waste from one source become the resources for other system by using the natural materials efficiently. Nature creates high quality structured materials with unique properties, that often very tough in artificial synthesis. Such natural materials are eco-friendly for long term ecological balance. Such complex designed materials attract researchers attention to study and mimic them to develop more and better sustainable materials.

Waxes are one of the most significantly adaptable lipid-derived materials, which is widely distributed across the domain of plants, animals, and microorganisms. These materials [4-6] have a huge demand in scientific and industry field due to their excellent chemical stability, water-repellent nature and protective properties. Nearly every biological kingdom utilizes natural waxes, which represents a diverse class of organic compounds which has an important evolutionary role. Historically [4] waxes has been a part of human civilization for a long period of time. Various natural waxes are produced in nature and one of the well documented and extensively studied wax is Bee wax [7]. The honey bee (*Apis mellifera*) is the most extensively documented producer of bee wax [8,9]. Bee wax is utilised for their nest construction as a result, the creation of unique hexagonal honeycomb structure. Due to its consistent chemical profile, honey bee wax has been standard in commercial and scientific research.

Stingless bees [10-12] are one of the most diverse groups of social insects found in tropical and subtropical regions which includes southeast Asia, Africa, South America Indian subcontinent etc. Due to their reduced and non-functional sting, they exhibit distinct nesting behaviour and nest architecture in the environment. Their hives (Figure 1a) are sophisticated biological fortresses and designed in such a way that they can withstand with tropical conditions like high humidity, high temperature etc. [13] The primary structural medium is a specialized organic mixture known as cerumen—a blend of wax secreted by abdominal gland of bees and botanical resins or propolis collected from the environment. For further strengthening of the bee hive, they create more protective layer, which is batumen, which is formed by the combination of cerumen and extra resin [14].

The construction of these hives demonstrates complex natural engineering [14,15]. The hives are generally constructed in tree cavities, rock cavities and underground spaces where they form there complex internal structures . The bees manipulate these materials to create a functional hierarchy, ranging from flexible involucrum sheets that provide thermoregulation for the brood to rigid storage pots for honey and pollen. The incorporation of plant produced resins into the wax gives this stingless bee nest material its unique mechanical strength, antimicrobial properties and chemical complexity which make them different from the honeybee wax structure. While bees wax has been studied extensively, the specific structural and chemical properties of stingless bee nest materials remain under-explored at molecular level. Most literature focus on the chemical markers of honey or the medicinal properties of propolis which are the major products of the beehive, leaving a void in the microstructural analysis of the hive itself. Without high-resolution characterization techniques, the understanding of the potential applications of these natural materials remains incomplete. In this article, a comprehensive analysis on structure and chemical composition of naturally prepared stingless bee hive (NP-SBH) was carried out using various advanced characterization techniques such as; XRD, FESEM, FTIR and HRTEM along with a debut of photoluminescence behaviour using PL spectroscopy.

## METHODOLOGY

The NP-SBH sample (Figure 1) was characterized using FESEM, HRTEM, FTIR, and XRD to study its morphology and functional groups present in it. The surface morphology and topographic features of the naturally aged beehive (Sample A+) were analysed using Field Emission Scanning Electron Microscopy (FESEM) [JEOL JSM-IT800]. The internal ultrastructure of the sample was observed at the nanometre scale using high resolution Transmission electron microscope (JEM-F200, JEOL). The degree of crystallinity and the phase composition of Sample were evaluated using X-Ray Diffractometer (XRD) [Rigaku Ultima-IV]. To determine the presence of functional groups, FTIR spectroscopy was employed. The spectra were recorded in the mid-infrared range (4000–1000 $cm^{-1}$). PL spectrometer was used to observe the emitted light from the material and the decay time was measured using time resolve mode.

## RESULTS AND DISSCUSSION

The surface morphology and microstructure of the NP-SBH sample using FESEM is presented in Figure 2. The micrographs revealed nanosheets and micro-oval structures throughout the sample with carbon richness as confirmed from EDAX (Figure S1). At higher magnification the oval shaped structure showed wrinkled structure on their surface, as a close resemblance to that of dry dates. SEM images at different magnification showed difference in their appearance like lower magnification showed the overall arrangement of particles in the matrix while higher magnification showed minute details about the micro particles and their surface texture (Figure 1). The average size of the micro-oval structure was estimated to be approximately 4-5 $\mu$m and distributed uniformly across the observed area (Figure2 c and d). The magnified pictures were taken at different levels ranging from 10 µm - 0.1 µm. The carbon Nano sheets of different sizes are also distributed throughout the sample along with the micro-oval structure, which has a close resemblance to stacking of graphene sheets. EDAX elemental analysis identified carbon as the dominant constituent in the sample while oxygen was found in low percentage in the sample (Figure S1).

In order to confirm the C-sheet found from FESEM data, HRTEM have been performed on the same sample, which is shown in figure-3. From this figure one can visualize the clear C-sheet in fig 3 (a and b) and the c-richness is confirmed from EDAX data as shown in fig3 (c). To further investigate the crystallinity of the sheet, the selected area electron diffraction (SAED) were carried out (Figure 3e), where ring like patterns are observed. From the associated ring patterns, the interplanar spacing (d) were calculated as; 3.39 Å, 2.14 Å and 1.23 Å for first three consecutive rings and the corresponding (h,k,l) planes are (0,0,2), (1,01) and (110) shown in figure 3(e). All these parameters obtained are well agreed with earlier reports [16-20] of graphene/graphite sheets. Further also we have obtained the high resolution lattice fringes (Figure 3 e), where the d value has been measured as 3.4 Å. It enhances the clarity of evidence in observing the graphene sheet in the NP-SBH sample. The scanning transmission electron microscopy (STEM) image has also been taken to confirm the richness of carbon contents, which is shown in figure 3d. The distribution of carbon atoms in Figure 3 (d) (blue) matches exactly as of the graphene sheet acquired in HRTEM image (Figure 3 a and b).

Further, XRD (Figure S2a) and FTIR (Figure S2b), have been performed for verifying the crystallinity and their respective functional groups. The clear peak at 2θ ~ 23.8° of XRD data corresponds to (002) plane, which confirm the presence of graphene structure in the sample [21,22]. The Peaks observed at around ~2916 $cm^{-1}$ and ~2848.24 $cm^{-1}$ were attributed to C–H stretching vibrations generally found in waxy materials [22,23]. 2916 $cm^{-1}$ corresponds to $CH_2$ symmetrical stretching while ~2848.24 $cm^{-1}$ corresponds to asymmetrical stretching. A strong peak near ~1707 $cm^{-1}$ suggested C=O stretching and a peak near ~ 1644 $cm^{-1}$ suggested C=C stretching. Additional peaks around 1462.30–1376.26 $cm^{-1}$ and 1034.74 $cm^{-1}$ correspond to CH bending and C–O stretching vibrations, respectively, confirming the presence of C-derived materials.

To debut some interesting physical properties of NP-SBH, we have started with photo luminescence spectroscopy, which is presented in Figure 4. From figure 4(a), it is clearly observed that, the emission of blue light correspond to the peak arises within 400 – 490 nm. The decay time of this blue emission was measured through time resolved photoluminescence mode, from which we found that, there exists two life times of 1.18 ns (42 %) and 5.41 ns (58 %) with average life time as 3.63 ns. The higher life time of 5.41 ns may be attributed to radiative recombination from the band edge and is responsible for the intrinsic blue emission. The lower life time of 1.18 ns could be due to non-radiative fast recombination through defects, trapping etc. As observed from the FESEM, the graphene sheets are embedded with C-rich micro-oval structures, so the quantum confinement may occur in sp2 domains of graphene due to surrounding sp3 matrix like earlier report on quantum dots [24-27]. In particular, the edge states of graphene sheets introduce localized electronic states within the band structure, which recombine radiatively. The quantum confinement may happen due to the wrinkleless over the surface of micro-oval structures (Figure 2d). There are also emission from associated defects of the materials, such as vacancies, dangling bonds, extra oxygen, where electrons get trapped. From our FESEM results such defects too corroborate the emission of blue light in the PL spectra. So quantum confinement in nano scale sp2 domains of graphene increases the effective band gap, which leads to blue emission in the materials. Such findings in the naturally derived materials may contribute easily for applications [28-34] in optoelectronic devices, bio imaging and sensing, flexible and wearable electronics etc. Another important observation in microscopic level is that, the sample is devoid of pores, which may prevent the outer environmental contents for applications [35-38] such as waterproof coating, food packaging, thermal sustainability etc. The green graphene sheets, may lead to eco-friendly bio-compatible engineering materials in the society, especially in health sector. Further investigations and discretization of various components of the materials in nano scale may reveal many hidden behaviour for participating in quantum technologies [39].

## CONCLUSIONS

By using FESEM and HRTEM, C-rich microoval structure and stacked graphene sheets were discovered from naturally produced stingless bee hive material (NP-SBH). The richness of Carbon contents was clearly verified from the EDAX experiments. The crystallinity of the graphene sheets was clarified from the selected area diffraction pattern rings and the HRTEM fringes. XRD and FTIR support the microscopic findings. Very interestingly, we observed the blue emission from the photo luminescence data with two decay times 1.18 ns (42 %) corresponds to the non-radiative recombination and 5.41 ns (58 %) correspond to the radiative recombination. The graphene sheets embedded in micro-oval structure matrix, the wrinkles over the micro-oval structures and the associated defects therein, may lead to quantum confinement, as a result of blue light emission. Both micro-structure and properties findings in stingless bee hive will certainly contribute in various technological applications for the development of society. Especially the discovery of green graphene will play an important role in green technologies for an eco-friendly environment.

## ACKNOWLEDGEMENTS

Director, CSIR-IMMT, Bhubaneswar, is highly acknowledged for his constant support and encouragements. Special acknowledgement to the characterization facilities of CSIR-IMMT, Bhubaneswar.

**FIGURES:**

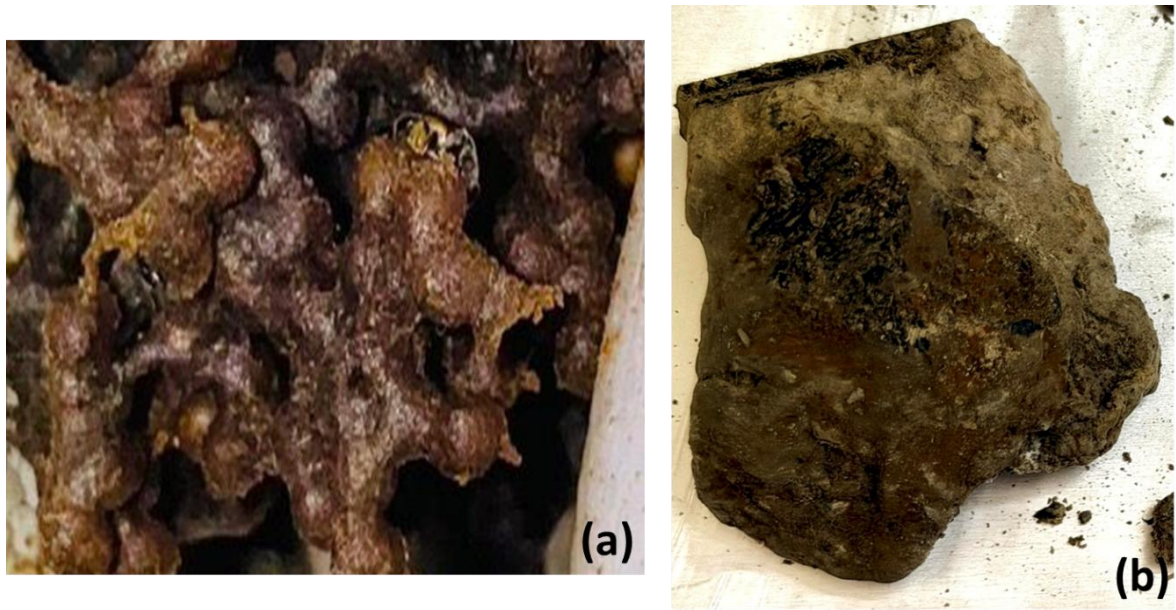

*Figure 1(a) Naturally produced stingless bee hive sample. (b) Dry solid sample of stingless bee hive.*

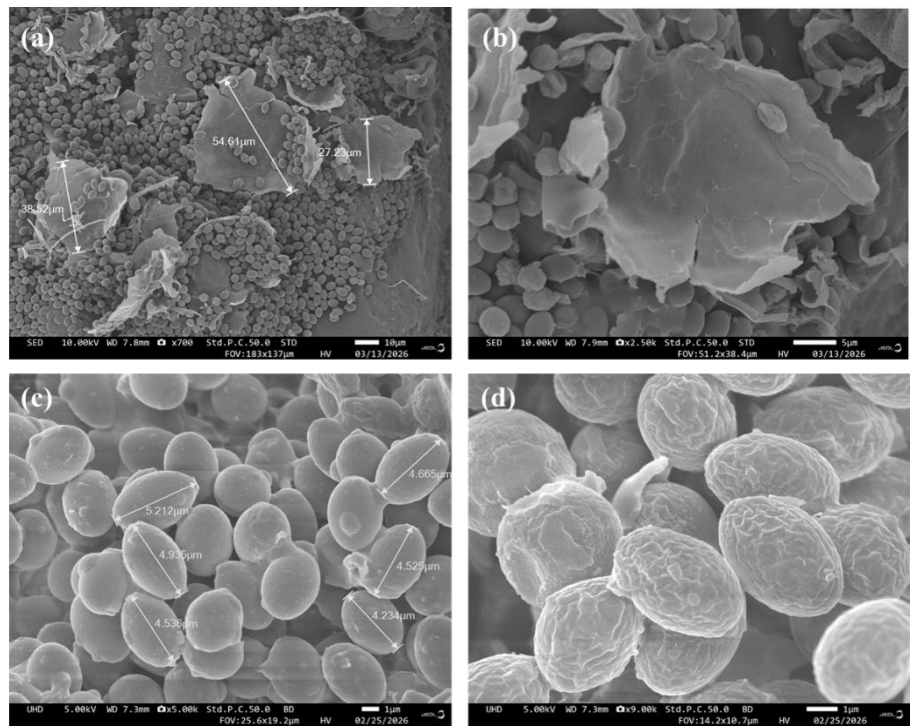

*Figure 2: FESEM images of NP-SBH sample at different regions with different magnifications*

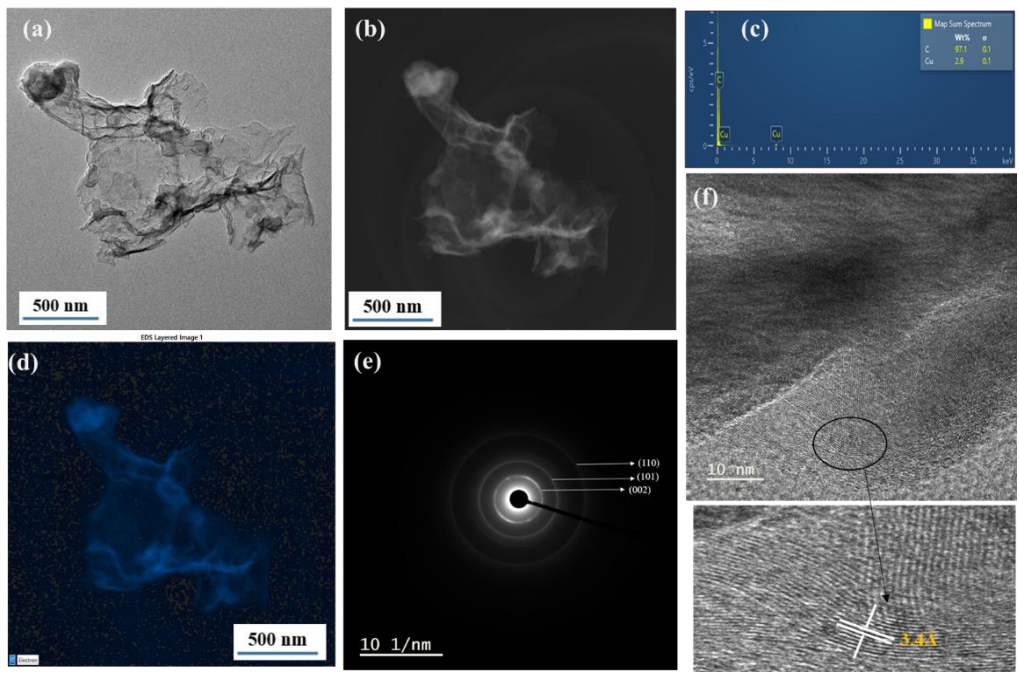

Figure 3 **(a)** TEM images of NP-SBH sample shows graphene sheets **(b)** The corresponding electronic image of graphene sheets in STEM mode **(c)** EDAX over graphene sheets shows the richness of Carbon atoms **(d)** Elemental mapping of graphene sheets shows the presence of Carbon atoms as the single blue colour **(e)** SAED pattern of the graphene sheets with their corresponding (h,k,l) planes **(f)** HRTEM images shows the interlayer spacing 'd' value.

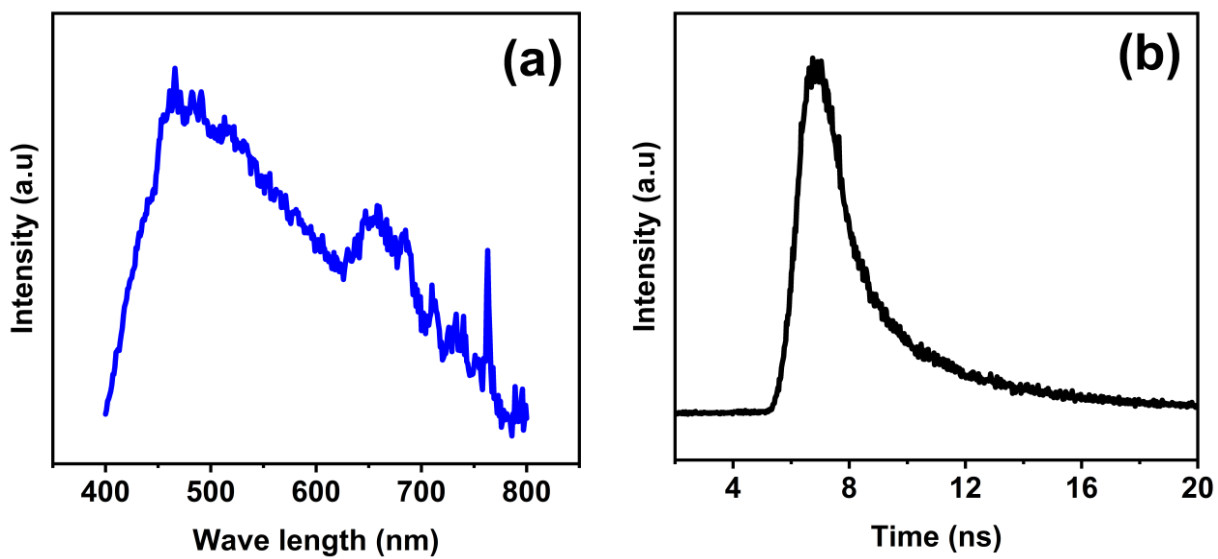

Figure 4**(a)** PL spectra of NP-SBH sample shows blue emission **(b)** Time resolved PL spectra of NP-SBH sample shows the life time of emission.

**SUPPLEMENTARY FIGURES:**

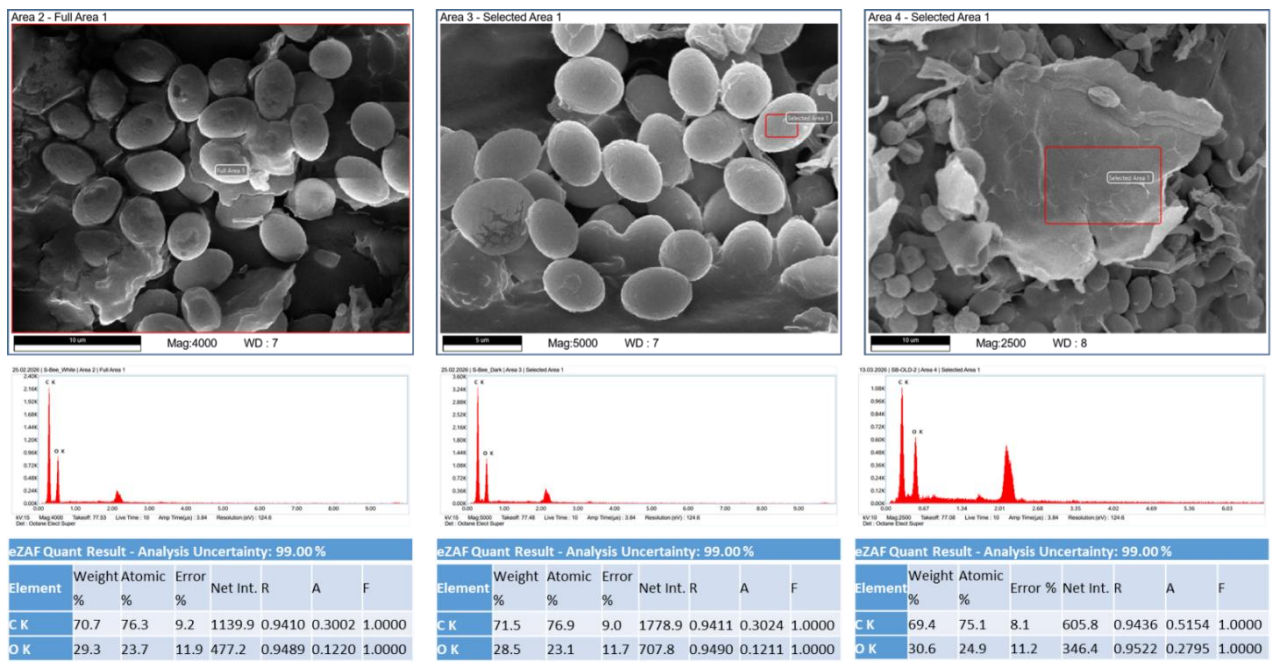

*Figure S1:* FESEM-EDAX of NP-SBH sample shows Carbon richness at different regions.

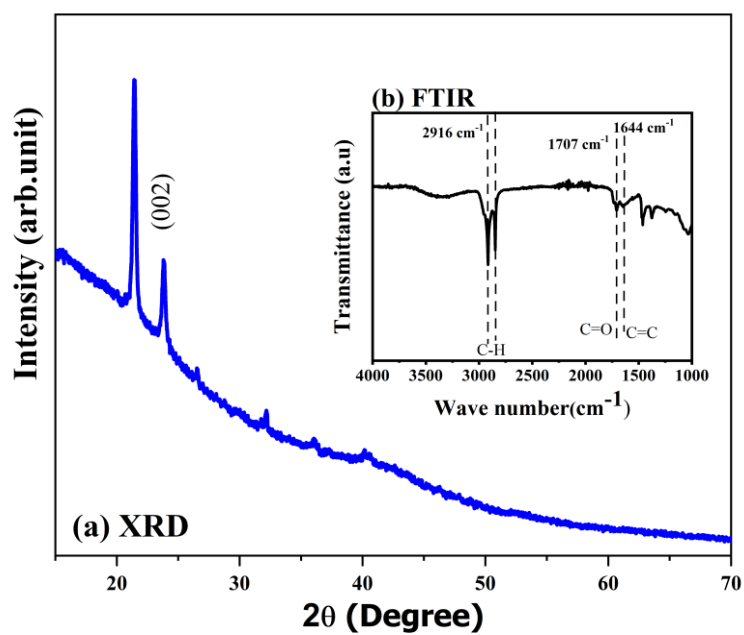

*Figure S2:* (a) XRD spectra of NP-SBH sample (b) FTIR spectra of NP-SBH sample.